\documentclass[12pt]{article}
\usepackage{amsmath}
\usepackage{epsf}
\usepackage{epsfig}
\usepackage{here}
\usepackage{amssymb}
\usepackage{citesort}
\usepackage{graphicx}
\usepackage{latexsym}
\usepackage{subfigure}
\usepackage{natbib}
\bibpunct{[}{]}{,}{}{}{} 
\newcommand{\be}{\begin{equation}}
\newcommand{\ee}{\end{equation}}
\newcommand{\bear}{\begin{eqnarray}}
\newcommand{\ear}{\end{eqnarray}}

\newsavebox{\LSIM}
\sbox{\LSIM}{\raisebox{-1ex}{$\ \stackrel{\textstyle<}{\sim}\ $}}

\newsavebox{\GSIM}
\sbox{\GSIM}{\raisebox{-1ex}{$\ \stackrel{\textstyle>}{\sim}\ $}}

\begin{document}
\begin{titlepage}
$\mbox{ }$
\vspace{.1cm}
\begin{center}
\vspace{.5cm}
{\bf\Large Suppressing Lepton Flavour Violation in a Soft-Wall Extra Dimension}\\[.3cm]
\vspace{1cm}
Michael Atkins$^{a,}$\footnote{m.atkins@sussex.ac.uk}
and 
Stephan J.~Huber$^{a,}$\footnote{s.huber@sussex.ac.uk} \\ 
\vspace{1cm} {\em  
$^a$Department of Physics \& Astronomy, University of Sussex, Brighton
BN1 9QH, UK }\\[.2cm]

\end{center}
\bigskip\noindent
\vspace{1.cm}
\begin{abstract}
A soft-wall warped extra dimension allows one to relax the tight constraints imposed by electroweak data in conventional Randall-Sundrum models. We investigate a setup, where the lepton flavour structure of the Standard Model is realised by split fermion locations. Bulk fermions with general locations are not analytically tractable in a soft-wall background, so we follow a numerical approach to perform the Kaluza-Klein reduction. Lepton flavour violation is induced by the exchange of Kaluza-Klein gauge bosons. We find that rates for processes such as muon-electron conversion are significantly reduced compared to hard-wall models, allowing for a Kaluza-Klein scale as low as 2 TeV. Accommodating small neutrino masses forces one to introduce a large hierarchy of scales into the model, making pressing the question of a suitable stabilisation mechanism.
\end{abstract}
\end{titlepage}
\section{Introduction}

Over the last ten years there has been a large increase in the study of extra dimensional models following the realisation that they could help explain some of the unresolved problems in the Standard Model (SM). In 1999, Randall and Sundrum showed that a warped extra dimension could offer a geometric solution to the gauge hierarchy problem \cite{Randall:1999ee}. In the original Randall-Sundrum (RS) model, the fifth dimension consists of a slice of AdS space bounded by ultraviolet (UV) and infrared (IR) branes. The warped space produces an exponential difference in energy scales between the two branes which solves the hierarchy problem. Matter fields were originally confined to the IR brane, however, it was soon realised that by allowing fermions to propagate in the extra dimension, the SM fermion mass hierarchy can be explained. By varying the location of the fermion wavefunctions in the fifth dimension, the full scale of fermion masses from neutrinos to the top quark can be generated using only order unity parameters \citep{Grossman:1999ra,Gherghetta:2000qt,Huber:2000ie}. This setup also contains a built in mechanism suppressing unobserved flavour changing processes that result from couplings between SM fermions and excited gauge bosons which appear in the model \citep{Gherghetta:2000qt,Huber:2003tu,Agashe:2004cp}.

Further interest in warped extra dimensions was generated by the AdS/CFT conjecture, when it was realised that the RS scenario is holographically dual to strongly coupled 4D field theories \citep{ArkaniHamed:2000ds,Rattazzi:2000hs,PerezVictoria:2001pa}. It was in this context, studying AdS/QCD models, that the idea of a soft wall was first introduced \citep{Karch:2006pv}. The soft wall is realised by removing the IR brane so the extra dimension extends to infinity, and replacing it with a smoothly varying spacetime cut off. The original AdS/QCD motivation for this was to more faithfully reproduce the linear Regge-like mass squared spectrum of excited mesons as opposed to the usual quadratic spectrum found in hard wall RS models.

Inspired by the possibility of qualitatively different phenomenology, the soft wall scenario was subsequently applied to modelling electroweak physics \citep{Falkowski:2008fz,Batell:2008me}. These models successfully showed that a soft wall extra dimension is generally less constrained by electroweak precision observables than its hard wall counterpart, typically allowing Kaluza Klein (KK) modes with masses of a few TeV. An important issue is related to the stability of a soft-wall setup, which is an open question in the models discussed in Refs.~\citep{Falkowski:2008fz,Batell:2008me}.
Such a mechanism was suggested in Ref.~\cite{Cabrer:2009we}, promising the soft-wall extra dimension to equally well resolve the gauge hierarchy problem. 

With the removal of the hard-wall brane the Standard Model matter fields must necessarily propagate in the bulk. Graviton fluctuations and gauge fields were successfully analysed in this background but it was found that fermions presented particular technical difficulties and only a simplified single generational model was developed. Later studies of fermions in a soft-wall extra dimension have developed solutions to the fermion problem \cite{Delgado:2009xb, MertAybat:2009mk, Gherghetta:2009qs} and have considered the experimental constraints imposed by the electroweak observables. However, the fermion flavour pattern of the SM has not been considered in much detail, in particular with respect to the generation of neutrino masses and the experimental bounds on lepton flavour violation.

In this paper we present a numerical solution to analyse a single generation of fermions in the soft-wall extra dimension. We extend this solution to three generations by treating flavour mixings as perturbations to the original solutions, and apply it to the lepton sector of the SM. We construct a setup, where the lepton flavour pattern is accommodated by flavour dependent localisations. It is shown that in order generate small Dirac neutrino masses by this mechanism we need to introduce a hierarchy of scales of order $10^{15}$ into the model, making crucial the issue of a suitable stabilisation mechanism. 
We finally carry out an analysis of the constraints coming from various lepton flavour violating processes, averaging over random order unity Yukawa couplings, and find that models with only a modest hierarchy of scales are relatively mildly constrained, whereas the model with a large hierarchy allowing sub-eV neutrino masses lies well within current experimental constraints, even for a KK scale\footnote{When referring to the soft wall model we take the KK scale to be the mass of the first KK mode of the Z boson.} of 2 TeV. In the latter, flavour violation is considerably suppressed relative to its hard wall counterparts, such as the ones analysed in \citep{Huber:2003tu,Moreau:2005kz}, and the range of masses lies in the reach of the LHC experiment.

At this stage we do not try to accommodate the flavour structure of the quark sector, which should be possible in a similar way. Also we reproduce the neutrino masses and mixings only at the qualitative level, which is sufficient to estimate the rates of lepton flavour violation.

\section{Bulk Fields in a Soft-Wall Extra Dimension}

Our conventions follow most closely those laid down in Refs.~\cite{Batell:2008me, Gherghetta:2009qs}. The 5D spacetime has metric
\begin{equation}\label{swmetric}ds^2 = e^{-2A(y)}\eta_{MN}dx^Mdx^N,\end{equation}
where $y$ represents the extra spatial dimension and $\eta_{MN}=\mbox{diag}(+,-,-,-,-)$. We take a pure AdS metric, $A(y)=\log ky$, where $k$ is the AdS curvature scale. There is no IR brane, the extra dimension extends to infinity and the soft wall is introduced via a dilaton field $\Phi$ with the action describing gauge and matter fields given by
\begin{equation}\label{action} S = \int d^4x \int_{y_0}^\infty dy \: \sqrt{g} e^{-\Phi} \mathcal{L}.\end{equation} 
Here $y_0=1/k$ is the location of the UV brane. The dilaton field is taken to have the following power law behaviour 
\begin{equation}\label{dilaton}\Phi(y) = (\mu y)^2 .\end{equation}
The dimensionful parameter $\mu$ will set the mass scale of the lightest KK excitations.
The behaviour of other powers has been discussed in detail in Ref. \cite{Delgado:2009xb}. It is shown in Ref. \cite{Batell:2008me} that an appropriate form for the Higgs VEV in such a background is given by
\begin{equation}\label{vev}h(y)=\eta k^{3/2} \mu^2 y^2 \end{equation}
where $\eta$ is a dimensionless $\mathcal{O}(1)$ coefficient.

\subsection{Massive Gauge Fields}

In Refs. \cite{Batell:2008me,Gherghetta:2009qs} only massless gauge fields are considered as the gauge couplings being considered are assumed to be between quarks and massless gluons. The flavour changing neutral currents (FCNC) we will be considering here are mediated by the massive Z boson and so we first develop the solutions for such a gauge field. A massive gauge field propagating in the bulk has the Lagrangian
$$  \mathcal{L} = -\frac14 F_{MN}F^{MN} + \frac12 M_A^2 A_M A^M .$$
The mass term $M_A^2$ for the weak gauge bosons arises from spontaneous symmetry breaking and with the Higgs VEV as given in (\ref{vev}) we have
$M_A^2=\frac12 \: g_5^2 \: h(y)^2.$
Varying the action (\ref{action}) we obtain the equation of motion
$$ \frac{1}{\sqrt{g}} \partial_M \left( \sqrt{g} \: g^{MN}g^{RS}F_{NS}\right) -g^{MN}g^{RS}F_{NS} \partial_R \Phi - M_A^2 G^{MN}A_M = 0.  $$
Imposing the gauge $A_y=0$, inserting the Kaluza Klein (KK) reduction,
\begin{equation}\label{KKgSW}A_\mu(x,y)=\sum_{n=0}^\infty A_\mu^{(n)}(x) f_A^{(n)}(y),\end{equation}
and requiring the $A_\mu^{(n)}(x)$ to be mass eigenstates, we find the $f_A^n$ have to satisfy
\begin{equation}\label{gaugeeom}\left( \partial_y^2 -\left(\frac1y + \Phi'\right)\partial_y -\frac{1}{(ky)^2}M_A^2 + m_n^2 \right)f_A^{(n)}(y)=0 \end{equation}
and are canonically normalised by
$$\int_{y_0}^\infty dy \: e^{-(A+\Phi)}f_A^{(m)}(y)f_A^{(n)}(y) = \delta^{mn}.$$
A complicated analytic solution of the above equation of motion was developed in \cite{Delgado:2009xb} but for our purposes we find it more convenient to solve the equation of motion numerically. We apply Neumann boundary conditions to the wavefunctions at $y_0$ and vary $\eta g_5$ in order to find a normalisable solution with the appropriate 4D zero mode mass of $m_0 \simeq 91$ GeV for the Z boson.

\begin{figure}[ht]
\centering
\subfigure[]{
\includegraphics[scale=0.55]{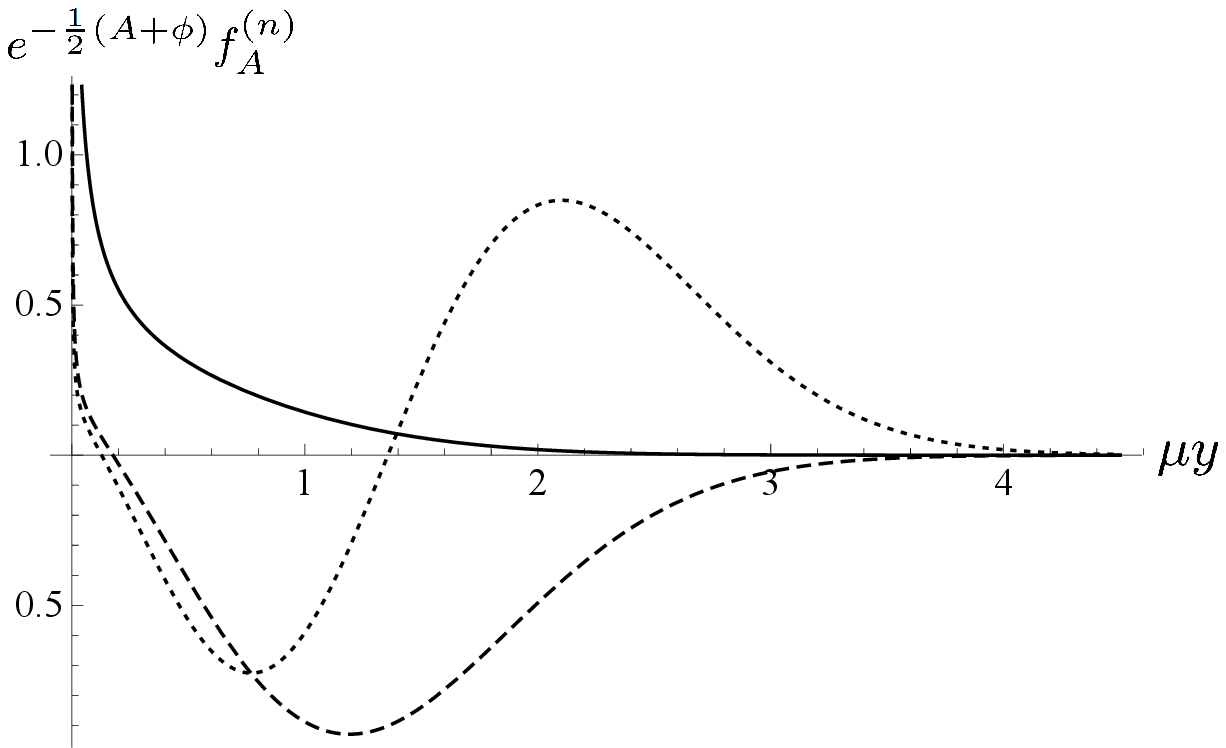}
\label{fig:subfig1}
}
\subfigure[]{
\includegraphics[scale=0.55]{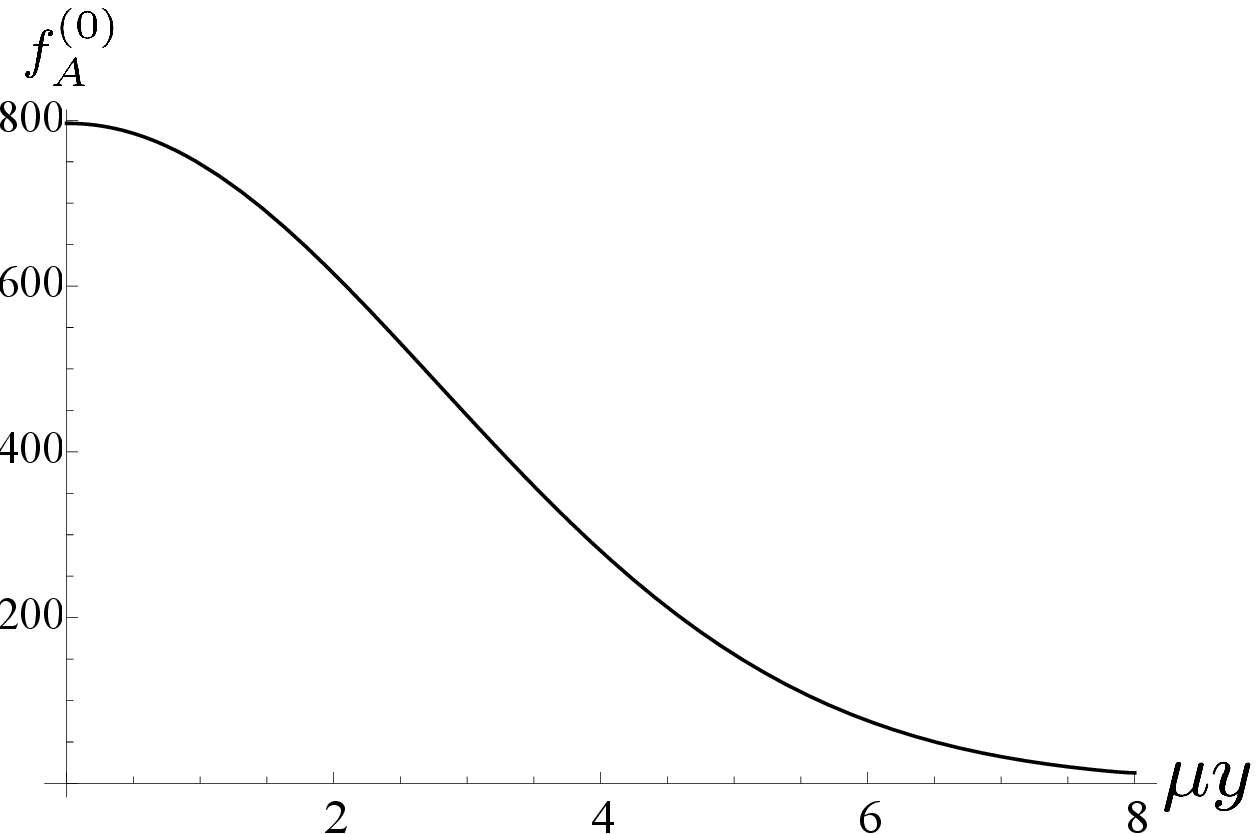}
\label{fig:subfig2}
}

\label{fig:gaugeprofiles}
\caption{Z boson profiles for $k=10^7$ TeV and $\mu=1$ TeV. (a) The zero mode (solid), first (dashed) and second (dotted) KK modes plotted with respect to a flat metric. (b) The zero mode profile as it couples to fermions.}
\end{figure}

It is interesting to note that unlike in hard-wall models, the profile of a massive gauge boson is independent of the curvature scale $k$. This can easily be seen by looking at the equation of motion (\ref{gaugeeom}) where $k$ only appears in the term involving the 5D mass $M_A$ which is the term that must be varied in order to generate the correct zero mode mass. The profile of the first few KK modes of the Z boson are plotted in Fig.~1. In Fig.~\ref{fig:subfig1} the zero mode and the first two KK modes are plotted with respect to a flat metric. Fig.~\ref{fig:subfig2} shows the profile of the zero mode in the form that it couples to the fermions (see later). The UV behaviour of the zero mode is less flat than in hard wall models and could possibly lead to large violations of universality of gauge couplings once fermions reside at different locations in the extra dimension. Note also that as is the case for massless gauge bosons \cite{Batell:2008me, Gherghetta:2009qs} the higher KK modes become more and more IR localised, a fact that will be important when considering couplings between higher gauge modes and fermions.

The KK spectrum for the Z boson for different values of $\mu$ with $k=10^7$ TeV is

\begin{tabular}{ l l l l}
& & & \\
$\mu=\frac12$ TeV: & $m_0=0.091$ TeV & $m_1=1.3$ TeV & $m_2=1.8$ TeV \\
$\mu=1$ TeV: & $m_0=0.091$ TeV & $m_1=2.2$ TeV & $m_2=3.0$ TeV \\
$\mu=2$ TeV: & $m_0=0.091$ TeV & $m_1=4.1$ TeV & $m_2=5.8$ TeV \\
$\mu=4$ TeV: & $m_0=0.091$ TeV & $m_1=8.2$ TeV & $m_2=12$ TeV\\
&&&
\end{tabular}

Note that the mass of the first KK mode $m_1 \sim 2\mu$, and hence the KK scale, $M_{KK}$, scales with $\mu$. Higher modes follow a Regge like spectrum $m_n^2 \sim n$.

\subsection{Fermions}

We consider 5D Dirac spinors $\Psi_L$ and $\Psi_R$ which are components of doublets and singlets under $SU(2)_L$ respectively. Note that $L$ and $R$ do not denote chiralities, but are related to the charges under $SU(2)_L$. 
 The chiral projections of these spinors are $\Psi_{L\pm}=\frac{1}{2}(1\mp\gamma^5)\Psi_L $, same for $\Psi_R$. The action for two free fermions in the bulk is
\begin{eqnarray*}S= \int d^4x \int_{y_0}^\infty dy \: \sqrt{g} e^{-\Phi} \left[ \frac{1}{2}\left( \: \overline\Psi_L i e_A^M \gamma^A D_M \Psi_L - D_M \overline\Psi_L i e_A^M \gamma^A \Psi_L\right)  - M_L \overline\Psi_L \Psi_L\right. \\
\left.+ \frac{1}{2}\left( \: \overline\Psi_R i e_A^M \gamma^A D_M \Psi_R - D_M \overline\Psi_R i e_A^M \gamma^A \Psi_R\right)  - M_R \overline\Psi_R \Psi_R\right]. \;\;\;\;\end{eqnarray*}
The f\"{u}nfbein and spin connection for the metric (\ref{swmetric}) are $e_A^M=e^{A(y)}\delta_A^M$ and $\omega_M=\left( -\frac{A'}{2}\gamma_\mu \gamma^5,0\right) $ and the covariant derivative is then $D_M=\partial_M + \omega_M$. $M_{L,R}$ are the 5D Dirac masses related to $\Psi_{L,R}$.

The difficulty of placing uncoupled fermions in the soft wall background is well documented in Ref. \cite{Batell:2008me}, all solutions suffer from divergent gauge couplings for high enough KK modes. The underlying problems stem from the non compact nature of the extra dimension and it is shown that in order to find workable normalisable solutions, the Yukawa couplings between fermions and the Higgs must be taken into account. An alternative approach to introducing Yukawa couplings was presented in Ref. \cite{MertAybat:2009mk} where a $y$ dependent Dirac mass term is introduced, somewhat like the $y$ dependent bulk mass arising from the Higgs VEV in the case of the massive gauge boson above. Here, however we will stick to constant Dirac mass terms $M_L$ and $M_R$ and introduce Yukawa couplings into the action:
$$S_{\rm Yuk}=-\int d^4x\int_{y_0}^\infty dy \sqrt{g} e^{-\Phi} \frac{\lambda_5}{\sqrt{k}}\left[\:\overline\Psi_L(x,y)h(y)\Psi_R(x,y) + \overline\Psi_R(x,y)h(y)\Psi_L(x,y) \right]. $$
Defining $\Psi_{L,R} = e^{2A + \Phi/2}\psi_{L,R}$ and $m(y)=\frac{\lambda_5}{\sqrt{k}}h(y),$ the equations of motion are
$$   i \gamma^{\mu}\partial_{\mu} \psi_{L,R\pm}  \pm \partial_y \psi_{L,R\mp}  - e^{-A}M_{L,R} \psi_{L,R\mp} -e^{-A}m(y) \psi_{R,L\mp}  =0. $$
Using the KK reduction
$$\psi_{L,R\pm}(x,y)=\sum^\infty_{n=0}\; \psi_\pm^{(n)}(x)\;f_{L,R\pm}^{(n)}(y),$$
and requiring the $\psi_\pm^{(n)}(x)$ to be mass eigenstates, the $f^{(n)} $s will be given by
\begin{equation}\label{sweom}\pm \partial_y\left(\begin{array}{c}f^{(n)}_{L\pm}\\ f^{(n)}_{R\pm}\end{array}\right)+e^{-A}\left(\begin{array}{cc} M_L& m(y)\\ m(y)&  M_R\end{array} \right)
\left(\begin{array}{c}f^{(n)}_{L\pm}\\ f^{(n)}_{R\pm}\end{array}\right)=m_n
\left(\begin{array}{c}f^{(n)}_{L\mp}\\ f^{(n)}_{R\mp}\end{array}\right).
\end{equation}
The $\psi_\pm^{(n)}$ will be canonically normalised by
$$
\int_{z_0}^{\infty}\;dy\; \left( f^{(m)}_{L\pm}\: f^{(n)}_{L\pm} \;+\; f^{(m)}_{R\pm}\: f^{(n)}_{R\pm}\right) =\delta_{mn}.
$$
We impose Dirichlet boundary conditions at $y_0$ for $f^{(n)}_{L-}$ and $f^{(n)}_{R+}$ in order to obtain a chiral 4D theory.

Analytic solutions of (\ref{sweom}) are only possible for a small set of Dirac mass terms, namely $M_L=M_R$  and $M_L+M_R \pm k = 0$ \cite{Gherghetta:2009qs}. As in hard-wall models, the Dirac mass parameters $M_{L,R}$ dictate how the fermion is localised in the extra dimension and it is convenient to parametrise these in terms of the AdS curvature, $M_{L,R}=c_{L,R}k$. Unfortunately the sets of parameters for which analytic solutions are available do not explore the full geography of possible mass parameters and as we shall see may lead to situations with unacceptably large rates of 
flavour violation. Ideally we would like to solve (\ref{sweom}) for any set of Dirac masses and this requires a numerical approach. 

The numerical solution we have developed involves a shooting type method. In order for the solutions to be normalisable they must not diverge in the IR and this only occurs for the correct choice of $m_n$, thus generating the KK spectrum. We choose a suitably large distance $L$ into the IR and solve the equations of motion subject to the UV boundary conditions and a starting choice for $m_n$. We then iterate the solution using Newton's method in order to find a value for $m_n$ such that two of the solutions e.g. $f^{(n)}_{L+}$ and $f^{(n)}_{R-}$ converge to zero at $y=L$. The equations of motion then automatically ensure that the other two solutions will also converge to zero for large $y$. Our solution has the advantage that it seems to be quite capable of finding solutions even for large values of the AdS curvature scale $k$, however it is not so suited to finding solutions for multiple generations of fermions as is done in Ref \cite{Gherghetta:2009qs}.

\section{Leptons}
\subsection{General Considerations}\label{sect.gc} 
Due to the presence of the extra KK states in extra dimensional models, couplings between SM particles and their KK excitations can potentially lead to conflict with experimental observations. In the SM, a tight set of constraints comes from the experimental bounds on flavour changing neutral currents (FCNC). In the hard-wall Randall Sundrum model these processes have been investigated and are shown to occur at rates that are dependent on the fermion locations. However, there are certain choices of fermion locations which provide almost universal gauge couplings and these almost universal gauge couplings are the source of the so called RS GIM mechanism which suppresses FCNC \cite{Gherghetta:2000qt,Huber:2003tu,Agashe:2004cp}.

It was found in Ref.~\cite{Batell:2008me} that for fermions in the soft-wall background, the analytic solution with $M_L=M_R$ can produce a large hierarchy of masses but only one of the fermion pair ($\psi_+$, $\psi_-$) could reside in an area of universal gauge couplings and it was thus assumed that dangerous rates of FCNC would be generated in such a situation. In fact this is one of the main motivations for finding a numerical solution to the fermion equations of motion, in the hope that one would be able to find fermion locations which can give a large mass hierarchy and yet simultaneously reside in an area of universal gauge couplings.

The gauge interaction between bulk gauge bosons and fermions is given by
$$S_{\rm{Gauge}}= g_5 \int d^4x \int_{y_0}^\infty dy \: \sqrt{g} e^{-\Phi} \left[\overline\Psi_L e_A^M \gamma^A A_M \Psi_L + \overline\Psi_R e_A^M \gamma^A A_M \Psi_R \right].$$
The couplings of $\psi^{(0)}_\pm$ to different KK gauge modes is then
\begin{equation}\label{gauge coup} g^n_\pm = g_5 \int_{y_0}^\infty dy \; f_A^{(n)} \left[ \left(f^{(0)}_{L\pm} \right)^2 + \left(f^{(0)}_{R\pm} \right)^2 \right]. \end{equation}

The dependence of the gauge couplings on the fermion locations $c_{L,R}$ is shown in Fig.~2. It can be seen that for $c_L>1/2$ the couplings become universal. In the case where $c_L=c_R$,  the couplings of one of the fermions would lie in the universal region the other would lie in the opposite part of the plot. However, with opposite Dirac masses, $c_L=-c_R$, we are able to place both fermions in a region of universal coupling at the same time and we would thus hope to suppress FCNC.

In hard wall models the origin of the regions of universal couplings is quite clear and derives from the profile of the gauge field wavefunctions which are flat in the UV. Hence if fermion profiles are relatively UV localised the gauge couplings will be universal. However, in the soft wall model, looking at the profile of the zero mode of the massive gauge boson in Fig. \ref{fig:subfig2} it is certainly not flat and one may wonder why we still find regions of universal gauge couplings. The explanation can be seen by considering the fermion profiles. Fig. \ref{fig:uvlocal} shows the fermion wavefunctions contributing to the gauge coupling of $\psi_+$ for $c_L = - c_R = 0.7$ which are locations that live in an area of universal gauge couplings. Whilst $f^{(0)}_{L+}$ is heavily UV localised we see that $f^{(0)}_{R+}$ is actually peaked into the IR which we would expect to contribute to non-universal couplings. However, when one considers the relative size of the contributions of each of these wavefunctions to the gauge coupling as given by Eq. \ref{gauge coup} we find $\int (f^{(0)}_{R+})^2/\int (f^{(0)}_{L+})^2 \sim 3 \times 10^{-4}$ i.e. almost the entire contribution to the gauge coupling comes from $f^{(0)}_{L+}$ which is heavily UV localised. Although the relative dominance of $f^{(0)}_{L+}$ is not that clear to see from Fig. \ref{fig:uvlocal}, it becomes obvious when one realises that it has a value at $\mu y_0$ of about 2000. Also, its extreme UV localisation can be seen by the fact that 99\%  of the area of $(f^{(0)}_{L+})^2$ lies in the region $\mu y < 0.01$. Hence the dominant contribution to the gauge coupling comes from a region in the extreme UV where the gauge profile is effectively given by its UV boundary value, thus producing universal couplings for fermions.

\begin{figure}[ht]
\centering
\subfigure[]{
\includegraphics[scale=0.75]{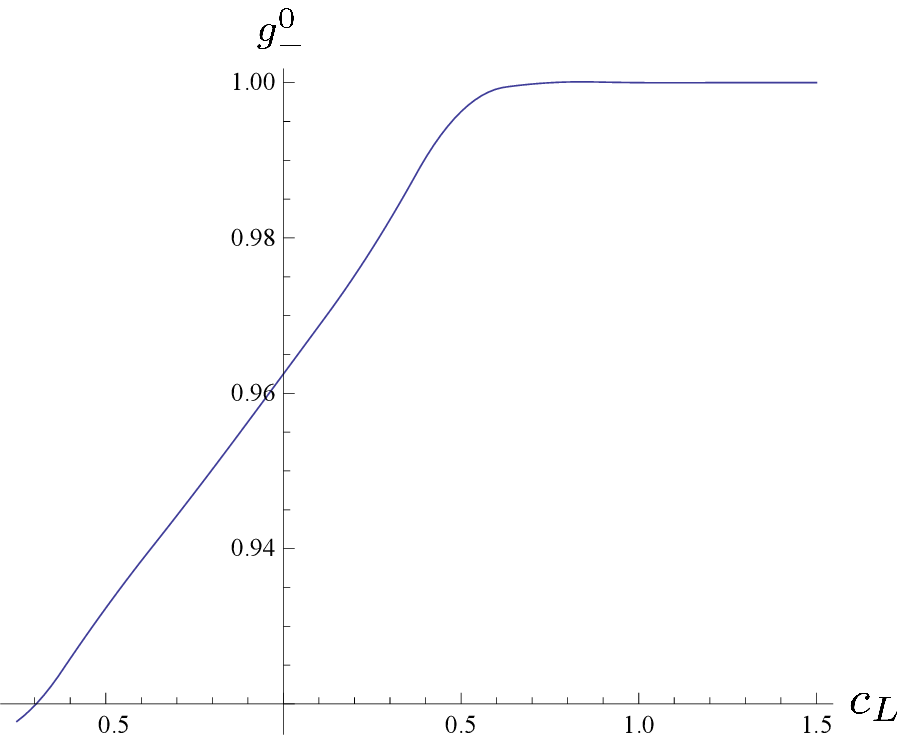}
\label{fig2:subfig1}
}
\subfigure[]{
\includegraphics[scale=0.75]{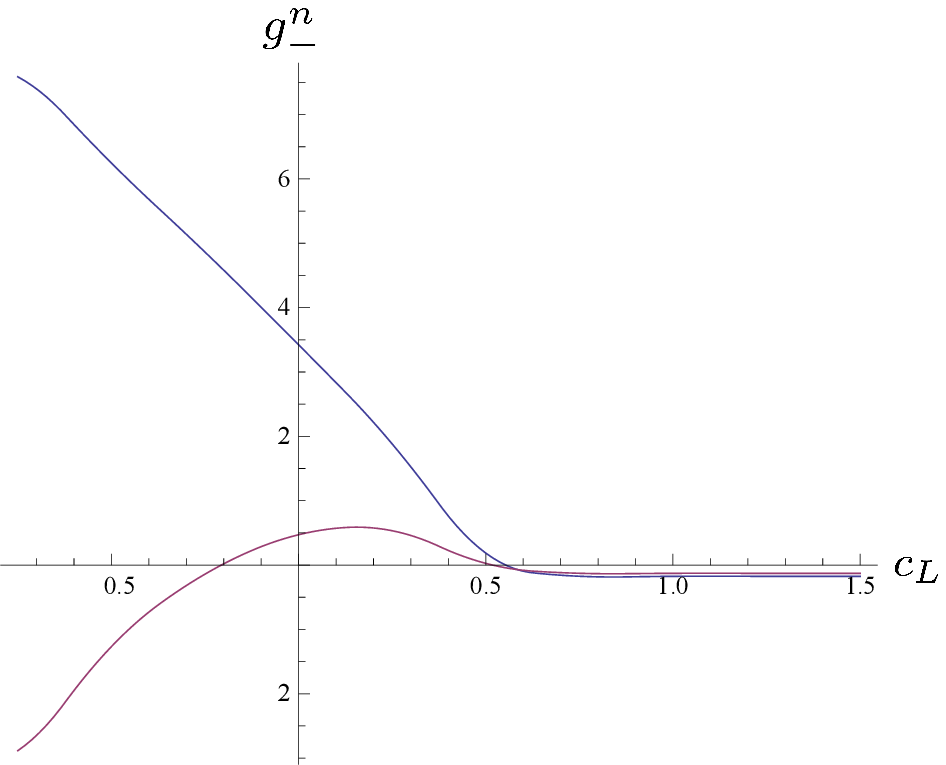}
\label{fig2:subfig2}
}
\label{fig:gaugecoup}
\caption{Gauge couplings of the Z boson (a) and its first two KK states (b) as a function of the fermion location, normalised so that the coupling to the zero mode is unity for large $c_L$. $k=10^7$ TeV, $\mu=1$ TeV and $c_L=-c_R$.}
\end{figure}

\begin{figure}
\centering
\includegraphics[scale=1]{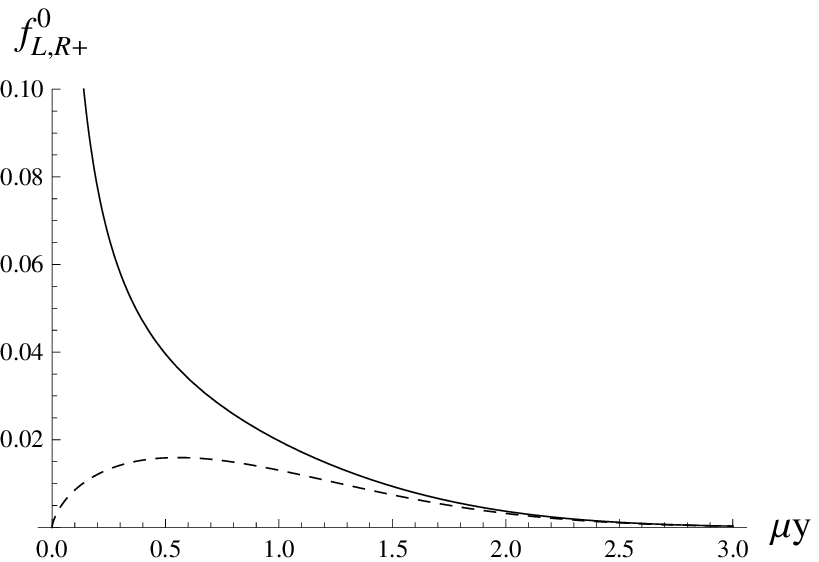}
\caption{Fermion zero modes with $c_L=-c_R=0.7$ for $k=10^7$ TeV and $\mu=1$ TeV. $f^{(0)}_{L+}$ solid, $f^{(0)}_{R+}$ dashed. Note that $f^{(0)}_{L+}$ takes a value of around 2000 at $y_0$.}
\label{fig:uvlocal}
\end{figure}

\begin{figure}
\centering
\includegraphics[scale=1]{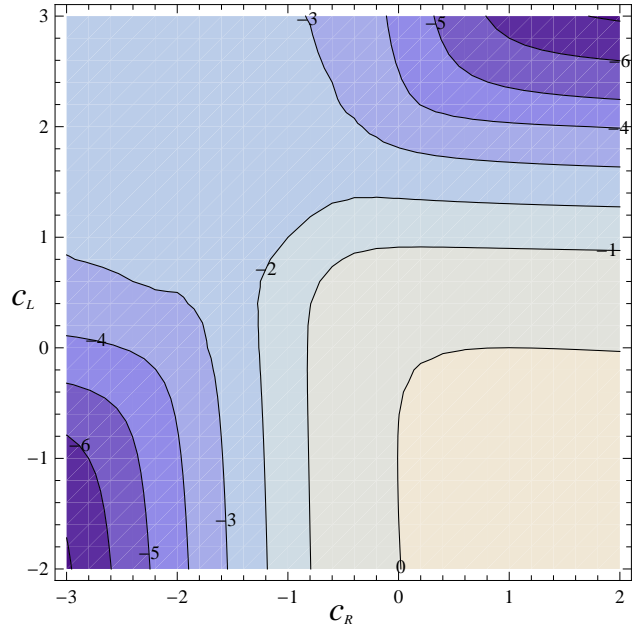}
\caption{Contour plot of $\log_{10}{(m_0/\mu)}$ for the zero mode masses of fermions with $\mu=1$ TeV and $k=10^3$ TeV.}\label{fig:contour}
\end{figure}

\begin{figure}
\centering
\includegraphics[scale=1]{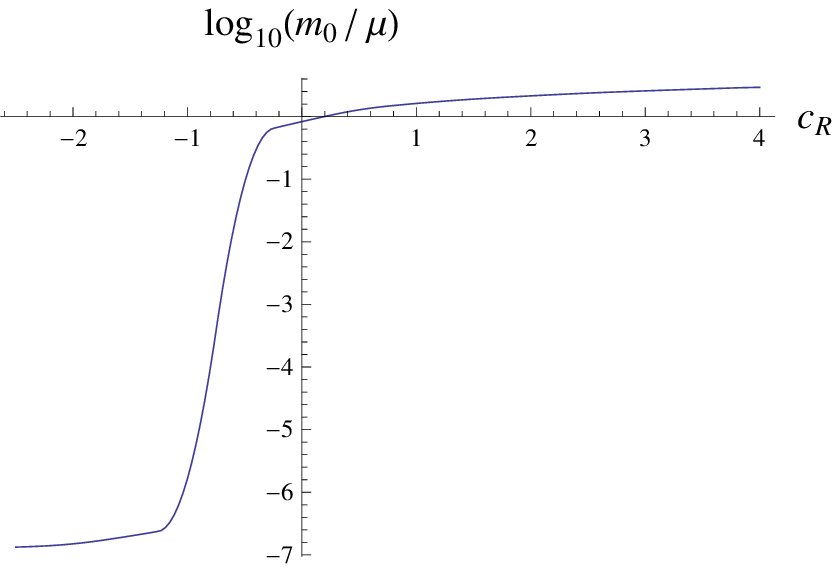}
\caption{Fermion zero mode masses with $c_L=-c_R$, $\mu=1$ TeV and $k=10^7$ TeV.}
\label{fig:crosssection}
\end{figure}

In order to generate the fermion mass hierarchy seen in the SM we have to carefully choose the $c$ parameters. The zero mode masses for different $c$ parameters can be seen in Fig. \ref{fig:contour}. Unfortunately, the shape of the plot presents a problem for simultaneously generating a large hierarchy of masses and universal gauge couplings. Whilst it is easy to generate a large hierarchy of masses for the choice of parameters $c_L=c_R$, as has been stated above, this is likely to lead to high rates of FCNC. In order to avoid these unacceptable rates we would like both the fermions to reside in an area of universal gauge couplings, this corresponds to the top left corner of the contour plot where we have $c_L>1/2$ and $c_R<-1/2$. In this area the zero mode mass bottoms out at around $\mu^2 / k$ and it is not possible to create a large mass hierarchy. The solution to this is to simply increase the hierarchy of scales in the model. Keeping $\mu=1$ TeV we can see from Fig. \ref{fig:crosssection} that with $k/ \mu = 10^{7}$ and $c_L=-c_R$ the zero mode masses could cover the full range of charged lepton masses whilst remaining in an area of universal gauge couplings. 

The reason for the zero mode mass having a minimum value for $c_L=-c_R$ can be seen by considering the two different ways small masses are generated in such models. The zero mode masses are generated via Yukawa couplings which involve the overlap between $\Psi_L$, $\Psi_R$ and the Higgs VEV. In the case where $c_L=c_R$, the wavefunctions of the zero mode fermions become oppositely localised and can be arranged to have an arbitrarily small overlap with each other thus creating arbitrarily small masses, this is the mechanism used in the ``split fermion'' model \cite{ArkaniHamed:1999dc}. However when we take $c_L=-c_R$ the fermions completely overlap each other and the zero mode mass is then entirely determined by their overlap with the Higgs. In RS models where the Higgs resides on the IR brane, this overlap can be made arbitrarily small by heavily localising both the fermions at the UV brane. However in the soft-wall model where the Higgs must necessarily propagate in the bulk and has a non zero value at the UV brane, the fermion wavefunctions will always have a minimum overlap with the Higgs at $y_0=1/k$ however much they are UV localised.

\subsection{Neutrino Masses}

We can generate small Dirac neutrino masses by introducing right-handed neutrinos and allowing Yukawa couplings between them, the Higgs and the left-handed neutrinos. A similar approach has been taken in the hard-wall case in Refs.~\citep{Grossman:1999ra,Huber:2001ug}. The left-handed neutrinos will share the same $c_L$ parameters as the corresponding left-handed charged leptons since they are part of the same doublet under $SU(2)_L$. We are then free to place the right-handed neutrinos  in a suitable location in order to generate sub-eV masses. However, since we still require $c_L > 1/2$ we again find that we are unable to generate such small masses without vastly increasing the overall hierarchy in the model. It seems necessary therefore to work with a hierarchy similar to that proposed in the original Randall Sundrum model. The issue of stabilising such a large hierarchy is an important question, and it would be very interesting to redo our analysis in context of the stabilised model proposed in Ref.~\cite{Cabrer:2009we}.
With this in mind we choose $k/\mu = 10^{15}$ and are able to produce neutrino masses of order $0.1$ eV by choosing $c_L=0.6$ and $c_R=-1.3$.

\subsection{Three Generations}

When incorporating all three generations of leptons into our model the Dirac mass terms $M_L$ and $M_R$ and the Yukawa coupling constants are promoted to $3\times3$ matrices mixing the different generations. We assume that the basis of states in which the  $M_L$ and $M_R$ are diagonal does not correspond to one in which the Yukawa couplings are diagonal. Rather than finding exact solutions for all three generations in such a scenario, our approach to this problem follows closely the method used in Ref.~\cite{Huber:2003tu} for the Randall Sundrum model. We solve the equations of motion individually for each generation with a Yukawa coupling $\lambda_5=1$, excluding fermion mixing, and use these solutions as basis from which we treat the full matrix of Yukawa couplings including mixings between the generations as perturbations. We specifically choose a large number of random Yukawa couplings, taking $\frac12 < |\lambda_{5ij}|<2$ with random sign\footnote{We do not consider CP violation, i.e.~we take $\lambda_{5ij}$ to be real.}, and require that the average zero mode masses reproduce the observed lepton masses. We also choose to locate the left-handed fields of each generation close to each other in order to generate large neutrino mixings, in the spirit of Ref.~\citep{Huber:2001ug}. However, we do not aim at reproducing the neutrino masses and mixings precisely. All we arrange for is an overall neutrino mass scale of order $0.1$ eV. Using our tools, a full model of neutrino masses could be constructed. However, for the following estimate of lepton flavour violation these details are not needed. Also we use the fact that neutrino mixings are order unity. 

In the case where we are not interested in generating neutrino masses via locations, we take only a moderate hierarchy of scales,  $k/ \mu = 10^{7}$. In this regime we choose the following three scenarios:

\begin{tabular}{ l l l l}
 & & &  \\
(A): & $c_{L1}=0.700$, & $c_{L2}=0.700$, & $ c_{L3}=0.700$, \\
   & $c_{R1}=-1.376$, & $c_{R2}=-0.903$, & $c_{R3}=-0.703$,  \\
   &  & & \\
(B): & $c_{L1}=0.720$, & $c_{L2}=0.700$, & $ c_{L3}=0.680$, \\
   & $c_{R1}=-1.373$, & $c_{R2}=-0.903$, & $c_{R3}=-0.704$,  \\
   &  & & \\
(C): & $c_{L1}=0.600$, & $c_{L2}=0.600$, & $ c_{L3}=0.600$, \\
   & $c_{R1}=-1.430$, & $c_{R2}=-0.980$, & $c_{R3}=-0.790$.  \\
 & & &  \\
\end{tabular}

In the regime where we can also generate neutrino masses, $k/ \mu = 10^{15}$ we choose:

\begin{tabular}{ l l l l}
 & & &  \\
(D): & $c_{L1}=0.60$, & $c_{L2}=0.60$, & $ c_{L3}=0.60$, \\
   & $c_{R1}=-0.82$, & $c_{R2}=-0.64$, & $c_{R3}=-0.55$.  \\
& & &  \\
\end{tabular}

Our choices for the different scenarios (A), (B) and (C) are to demonstrate the effects of degenerate $c_L$ localisation (A), small separation in the $c_L$ to introduce some non-universality in the left handed sector (B), and placing the left handed fermions closer to the IR brane (C). We expect the behaviour to be quite general and thus only choose one scenario with a larger hierarchy. The mass of the first fermion KK states is about $1.5$ TeV.

\section{Flavour Violation}

With a full three generations of leptons implemented as above, the transformation to fermion mass eigenstates will induce flavour violating couplings to gauge fields, in particular the Z boson and its KK excitations\footnote{Note that we work in a gauge field basis, where the Z boson zero mode is massive. Its properly weighted wave function is not flat, resulting in non-universal couplings to fermions. Alternatively, one could work in a basis, where the zero mode is massless.  Then only the KK states would couple non-universally.}  
We define the neutral current gauge couplings in the basis of mass eigenstates as
$$ \mathcal{B}_{\pm}^{(n)} = \mathcal{U}_{\pm} \mathcal{G}_{\pm}^{(n)} \mathcal{U}_{\pm}^\dagger,$$
where the unitary matrices $\mathcal{U}_{\pm}$ diagonalise the full fermion mass matrices and $\mathcal{G}_{\pm}^{(n)}$ are diagonal matrices that contain the couplings of the $n$th KK state of the Z boson to each fermion state as derived from Eq.~\ref{gauge coup} and normalised to the coupling of the muon (see also Ref.~\cite{Huber:2003tu}). Flavour violation induced by these couplings is dependent on the non-universality in the couplings of different flavour states and the mixing between the states. Different fermion locations increase the non-universality but at the same time lead to small mixing angles. Conversely similar fermion locations produce large mixing but this is compensated by universal couplings.

As was done in Ref.~\cite{Huber:2003tu} we calculate the rates of the various flavour violating processes using the techniques developed for family non-universal Z' bosons \cite{Langacker:2000ju}. The main difference being that there is no mixing between the different KK states of the Z boson, while the zero mode also has flavour violating couplings.

The first process we consider is the tree level exchange of a Z boson and its KK states mediating the process $l_j \rightarrow l_i
l_i\bar{l}_i$. The rate for this process is given by \cite{Langacker:2000ju}
$$ \Gamma(l_j \rightarrow l_i l_i \bar{l}_i)=\frac{G_F^2m_{l_j}^5}{48 \pi^3}\left(2|C^+_{ij}|^2 + 2|C^-_{ij}|^2 + |D^+_{ij}|^2 + |D^-_{ij}|^2 \right)$$
where
$$ C^\pm_{ij}= \sum_n \frac{M_0^2}{M_n^2} (\mathcal{B}_{\pm}^{(n)})_{ij}(\mathcal{B}_{\pm}^{(n)})_{ii}$$
$$ D^\pm_{ij}= \sum_n \frac{M_0^2}{M_n^2} (\mathcal{B}_{\pm}^{(n)})_{ij}(\mathcal{B}_{\mp}^{(n)})_{ii}$$
where we take the sum over the Z boson zero mode and the first two KK modes. $M_n$ is the mass of the $n$th KK mode of the Z boson. Due to the Regge type behaviour of the KK spectrum it is not clear that the above series should converge. However as noted in Ref. \cite{Gherghetta:2009qs}, due to the increasing IR localisation of higher gauge boson KK modes, the couplings rapidly decrease and the series converges after only a few terms. In the fermion sector we take into account only the zero modes. Mixing with KK fermions is small, leading to negligible effects at the current precision.

The branching ratios for the above processes in the different scenarios we consider are then found to be

\begin{tabular}{ l c c c c}
& & & &\\ 
& (A) & (B) & (C) & (D) \\
${\rm Br}(\mu \rightarrow e e \bar{e}):$ & $ 2.7 \times 10^{-14} $ & $ 5.1 \times 10^{-12} $ & $ 4.6 \times 10^{-12}$ & $  2.5 \times 10^{-15}$\\
${\rm Br}(\tau \rightarrow \mu \mu \bar{\mu}): $ & $2.4 \times 10^{-14}  $ & $ 1.8 \times 10^{-12} $ & $ 7.0 \times 10^{-13} $ & $ 2.7 \times 10^{-12}$\\
${\rm Br}(\tau \rightarrow ee\bar{e}): $ & $ 2.6 \times 10^{-15} $ & $ 1.5 \times 10^{-12} $ & $ 6.6 \times 10^{-13} $ & $  2.8 \times 10^{-16} $\\
& & & &\\
\end{tabular}

\noindent
These numbers are obtained for a KK scale of 2 TeV. They are the result of averaging over random Yukawa couplings in the range stated above.
The experimental bound ${\rm Br}(\mu \rightarrow e e \bar{e})< 1.0 \times 10^{-12}$ \cite{Bellgardt:1987du} is satisfied in the cases (A) and (D). However, it appears that the couplings are not universal enough to allow for much separation between the left handed states (B), and placing the fermions too close to the IR brane (C) also exceeds the experimental bound. However, like in the hard-wall case, the rate for this process depends on the KK scale as $1/M_{KK}^4$. Thus with a KK scale of twice as big (i.e.~take $\mu=2$ TeV while keeping $k=10^{7}$ TeV) scenarios (B) and (C) would also acquire an acceptable rate. The experimental bounds for the other two processes ${\rm Br}(\tau \rightarrow \mu \mu \bar{\mu})< 2.1 \times 10^{-8}$ and ${\rm Br}(\tau \rightarrow ee\bar{e})< 2.7 \times 10^{-8}$ \cite{Hayasaka:2010np} are well satisfied in all the scenarios. Note that in Ref.~\cite{Huber:2003tu} in a case similar to (A), (C) and (D) a branching ratio ${\rm Br}(\mu \rightarrow e e \bar{e})=5\times10^{-14}$ has been found for a KK scale of 10 TeV, translating into ${\rm Br}(\mu \rightarrow e e \bar{e})=3\times10^{-11}$ for a KK scale of 2 TeV. This demonstrates that lepton flavour violation is suppressed by up to four additional orders of magnitude in the soft-wall case.

We are also able to calculate the expected rate of $\mu \rightarrow e$ conversion in a muonic atom. The most stringent bound comes from the Sindrum-II Collaboration \cite{Wintz:1998} in $^{48}_{22}\mbox{Ti}$ where ${\rm Br}(\mu^{-}N \rightarrow e^{-}N)<6.1\times10^{-13}$. We can calculate the branching ratio for this process by \cite{Langacker:2000ju}
$${\rm Br}(\mu^{-}N \rightarrow e^{-}N) = \frac{G_F^2 \alpha^3 m_{\mu}^5}{2 \pi^2 \Gamma_{\rm CAPT}}\frac{Z_{\rm eff}^4}{Z}|F_P|^2 \left(\left|B^-\right|^2+\left|B^+\right|^2\right)$$ 
where
$$B^\pm=\sum_n\frac{M_0^2}{M_n^2}\:(\mathcal{B}_{\pm}^{(n)})_{12}\:\left[(2Z+N)(B_{u_L}^{(n)}+B_{u_R}^{(n)})+(Z+2N)(B_{d_L}^{(n)}+B_{d_R}^{(n)})\right]^2$$
and we take 
$$B_{\psi_{L,R}}^{(0)}=g_{\psi_{L,R}},\;\;\;B_{\psi_{L,R}}^{(1)}=0.19\: g_{\psi_{L,R}}\;\;\;\mbox{and}\;\;\;B_{\psi_{L,R}}^{(2)}=0.14 \:g_{\psi_{L,R}}.$$
Here, $g_{\psi_{L,R}}$ are the usual Standard Model quark couplings and we have taken the approximate values of the quark couplings to the higher KK gauge modes from the values derived in Ref. \cite{Gherghetta:2009qs}. We also take $Z_{\rm eff}=17.6$, $F_P=0.54$ and $\Gamma_{\rm CAPT}=2.59 \times 10^6 \: \rm{s}^{-1}$ \cite{Bernabeu:1993ta}, where $Z_{\rm eff}$ is an effective atomic charge, $F_P$ is a nuclear matrix element and $\Gamma_{\rm CAPT}$ is the muon capture rate.

We find branching ratios for the different scenarios of

\begin{tabular}{ l c c c c}
& & & &\\
 & (A) & (B) & (C) & (D) \\
${\rm Br}(\mu^{-}N \rightarrow e^{-}N):$ & $ 1.6 \times 10^{-13} $ & $ 3.3 \times 10^{-11} $ & $ 2.7 \times 10^{-11}$ & $  1.4 \times 10^{-14}$\\
& & & &\\
\end{tabular}

Again we find scenarios (A) and (D) lie within the experimental bounds but separating the states (B) or placing them too close to the IR brane (C) produces an unacceptable rate. Again we find flavour violation suppressed with respect to the hard-wall case \cite{Huber:2003tu}, making a KK scale of 2 TeV consistent with observations. However, next-generation experiments, such as PRISM at JPARC with a reach of ${\rm Br}(\mu^{-}N \rightarrow e^{-}N)\sim 10^{-16}-10^{-18}$ could probe a KK scale of 6 - 20 TeV, i.e.~the interesting parameter range of the present model. 

A third set of processes considered in Ref. \cite{Langacker:2000ju} are one-loop radiative lepton decays. Here the decay width is
$$\Gamma(l_j\rightarrow l_i \gamma)=\frac{\alpha G_F^2 m_{l_j}^3}{8 \pi^4}\left(\left|\xi_+^{ij}\right|^2+\left|\xi_-^{ij}\right|^2\right),$$
where the dipole moment couplings of an on-shell photon to the chiral lepton currents are given by
$$\xi_\pm^{ij}=\sum_n\frac{M_0^2}{M_n^2}\left(\mathcal{B}_\mp^{(n)} m_l \mathcal{B}_\pm^{(n)} \right)_{ij},$$
where $m_l$ is the charged lepton mass matrix. Using this we obtain the following rates for radiative decays

\begin{tabular}{ l c c c c}
 & (A) & (B) & (C) & (D) \\
${\rm Br}(\mu \rightarrow e\gamma):$ & $ 2.0 \times 10^{-17} $ & $ 3.7 \times 10^{-15} $ & $ 3.4 \times 10^{-15}$ & $  1.2 \times 10^{-18}$\\
${\rm Br}(\tau \rightarrow \mu \gamma): $ & $5.2 \times 10^{-16}  $ & $ 1.9 \times 10^{-14} $ & $ 9.4 \times 10^{-15} $ & $ 4.1 \times 10^{-14}$\\
${\rm Br}(\tau \rightarrow e \gamma): $ & $ 3.7 \times 10^{-17} $ & $ 1.5 \times 10^{-14} $ & $ 9.2 \times 10^{-15} $ & $  2.6 \times 10^{-18}$\\
& & & &\\
\end{tabular}

All of these branching ratios lie well within the experimental bounds ${\rm Br}(\mu \rightarrow e\gamma)=1.2\times10^{-11}$ \cite{Edwards:1996te}, ${\rm Br}(\tau \rightarrow \mu\gamma)=4.4\times10^{-8}$ and ${\rm Br}(\tau \rightarrow e\gamma)=3.3\times10^{-8}$ \cite{:2009tk}. Again these rates are suppressed relative to their hard-wall counterparts.

 In scenario (D), processes such as $\mu \rightarrow e \gamma$ can also be mediated by the KK states of the sterile neutrinos. This process was investigated for the Randall-Sundrum model in Ref.~\cite{Kitano:2000wr}. We use the formalism developed there and we find that in our model the branching ratio for this process is given by the relative coupling strength of the muon, the W and the KK muon neutrino to the zero mode muon neutrino times a loop factor. Because of the large mass differences of the sterile KK neutrinos, the GIM mechanism breaks down. We assume that the neutrino mixing angles are large, not leading to any suppression of the rate. 
Given a relative coupling of $0.0057$ we find $\rm{Br}(\mu \rightarrow e\gamma)=1.5\times10^{-13}$ which again lies within the experimental bounds for a KK scale of 2 TeV.  

There are also contributions to $\mu \rightarrow e \gamma$ related to the exchange of KK fermions \citep{Agashe:2006iy,Perez:2008ee}, which were neither included in our estimate above nor in Ref.~\cite{Kitano:2000wr}. These contributions dominate the rate of radiative lepton decays in hard-wall models. A similar behaviour is likely in the soft-wall model. However, given the suppressed rates for flavour violation in the latter, we expect that even including these extra contributions, the rate for $\mu \rightarrow e \gamma$ and a KK scale of 2 TeV should not exceed the experimental bound. 

We have shown that lepton flavour violation can be suppressed in the soft wall model and the amount of suppression depends on the fermion locations and the hierarchy of scales, $k/\mu$. The reason for this suppression comes from the ability to place the fermions in regions of universal gauge couplings i.e. heavily UV localised as explained in Section \ref{sect.gc}. When the hierarchy of scales in the model is increased by fixing $\mu$ and increasing $k$ the gauge boson profiles remain unchanged whilst the fermion profiles become even more UV localised thus providing an even greater suppression of flavour violation.

\section{Conclusions}
In this paper we have studied the lepton sector of the SM in a soft-wall
extra dimension, applying flavour dependent fermion locations to
accommodate the observed lepton flavour structure. The Higgs is a bulk
field, with a VEV that increases near the soft wall. We have in particular
considered the inclusion of small Dirac neutrino masses and investigated
the constraints on the model from lepton flavour violation mediated by the
Z boson and its KK states. In order to do so we first developed solutions
for a massive gauge boson in the soft-wall background and found the profile
is independent of the AdS curvature scale.  In order to generate the masses
of the charged leptons whilst keeping the fermions located in an area of
almost universal gauge couplings we find that we need to increase the
hierarchy of scales in the model to around $k /\mu  = 10^{7}$. When
incorporating sub-eV neutrino masses we need a much larger hierarchy, and we
choose $k/\mu = 10^{15}$, similar to the hierarchy between the Planck and
the electroweak scales.

To incorporate three generations of leptons into our model we solve the
fermion equations of motion numerically, including an order one flavour
diagonal Yukawa coupling and use these solutions as a basis of states from
which we treat off-diagonal Yukawa couplings, connecting different
generations, as perturbations. The mass term related  to the diagonal
Yukawa coupling is necessary to generate a normalisable wavefunction and
cannot be treated as a perturbation. We can construct the full lepton mass
matrices, including KK states and diagonalise them to find the fermion
masses and mixings. However, to our level of precision we can neglect the
fermionic KK states. 
 The locations of the left-handed fermions are dictated by the fact that
we require large mixings in the neutrino sector.  We take a large number of
random Yukawa couplings and choose the locations of the right-handed
fermions so that the averaged zero mode masses reproduce the SM charged
lepton masses.

With the inclusion of off-diagonal Yukawa couplings, the transformation to
mass eigenstates produces flavour violating couplings. We calculated the
expected rates for various flavour changing processes for a number of
different scenarios. We found that the soft-wall model is in fact
mildly constrained when we consider a scenario with a low hierarchy of
scales such as $k /\mu  = 10^{7}$. The most stringent constraint comes from
$\mu \to e$ conversion in a muonic atom where we find that only the
scenario where all the left-handed leptons have degenerate locations well
toward the UV brane would occur at acceptable rates with a KK scale of 2 TeV.
This is a considerable suppression of lepton flavour violation compared to
hard-wall models, such as the one studied in Ref.~\cite{Huber:2003tu}.
Including a larger hierarchy of scales ($k /\mu  = 10^{15}$), it is also
possible to generate sub-eV Dirac neutrino masses. In this case the model
is even less constrained and most of the FCNC processes would occur at rates
well below the experimental bounds.
The most stringent bounds are coming from radiative decays, such as $\mu
\rightarrow e \gamma$. Again a KK scale of 2 TeV seems sufficient to keep
the rate below the experimental bound. Our estimate for this rate does not
include contributions from KK gauge bosons, and it would be interesting to
include these in a more detailed analysis. Another obvious direction of research would
be to extend the present setup to the quark sector, similar to an analysis that was
performed recently in much detail for the hard-wall model in Ref.~\cite{Bauer:2009cf}.

The soft-wall extra dimension continues to offer a valid model for
electroweak physics, with constraints from precision data relaxed compared
to the hard-wall model.
Having said this, we have found that with a (gauge boson) KK scale of 2 TeV the complete
lepton flavour structure can be accommodated while keeping rare processes
below experimental bounds. In our setup the KK states of fermions have
masses around 1.5 TeV, within reach of the LHC experiment. Thus the
soft-wall framework seems to offer an alternative when it comes to suppressing flavour violation to models relying on
flavour symmetries \cite{Perez:2008ee,Csaki:2008qq,delAguila:2010vg}, a bulk Higgs \cite{Agashe:2008fe} or to  utilising non-minimal representations under the $SU(2)_R$ bulk gauge symmetry \cite{Agashe:2009tu}. 
 
 The parameter range with a large hierarchy $k /\mu  = 10^{15}$ is both
attractive to further suppress flavour violation and necessary to
accommodate neutrino masses. This rises the important question whether such
a hierarchy can be stabilised, like in the way proposed in Ref.~\cite{Cabrer:2009we}. It
would be very interesting to extend our analysis to such a framework.

\section*{Acknowledgements}
We would like to thank Sebastian J\"ager for useful discussions.

\bibliography{bib}

\end{document}